\documentclass[conference]{IEEEtran}
\IEEEoverridecommandlockouts
\usepackage{cite}
\usepackage{amsmath,amssymb,amsfonts}
\usepackage{algorithmic}
\usepackage[none]{hyphenat}
\usepackage{graphicx}
\usepackage{textcomp}
\usepackage[dvipsnames]{xcolor}
\usepackage{xspace}
\usepackage{hyperref}
\usepackage{pifont}
\usepackage{arydshln}
\usepackage{enumitem}

\newcommand*{\ie}{i.e.,\@\xspace}

\newcommand*{\iot}{IoT\@\xspace}
    
\newcommand*{\etal}{\emph{et~al.}\@\xspace}

\def\BibTeX{{\rm B\kern-.05em{\sc i\kern-.025em b}\kern-.08em
    T\kern-.1667em\lower.7ex\hbox{E}\kern-.125emX}}
\begin{document}
\makeatletter
\newcommand{\linebreakand}{%
  \end{@IEEEauthorhalign}
  \hfill\mbox{}\par
  \mbox{}\hfill\begin{@IEEEauthorhalign}
}
\makeatother
\title{Supporting Early-Safety Analysis of IoT Systems by Exploiting Testing Techniques}

\author{\IEEEauthorblockN{Diego Clerissi}
\IEEEauthorblockA{\textit{Universiy of Milano-Bicocca} \\
Milano, Italy \\
diego.clerissi@unimib.it}
\and
\IEEEauthorblockN{Juri Di Rocco}
\IEEEauthorblockA{\textit{University of l'Aquila} \\
L'Aquila, Italy \\
juri.dirocco@univaq.it}
\and
\IEEEauthorblockN{Davide Di Ruscio}
\IEEEauthorblockA{\textit{University of l'Aquila} \\
L'Aquila, Italy \\
davide.diruscio@univaq.it}
\and
\IEEEauthorblockN{Claudio Di Sipio}
\IEEEauthorblockA{\textit{University of l'Aquila} \\
L'Aquila, Italy\\
claudio.disipio@univaq.it}
\and
\IEEEauthorblockN{Felicien Ihirwe} 
\IEEEauthorblockA{\textit{University of l'Aquila} \\
L'Aquila, Italy \\
jeanfelicien.ihirwe@graduate.univaq.it}
\and
\IEEEauthorblockN{Leonardo Mariani}
\IEEEauthorblockA{\textit{Universiy of Milano-Bicocca} \\
Milano, Italy \\
leonardo.mariani@unimib.it}
\and 
\IEEEauthorblockN{Daniela Micucci}
\IEEEauthorblockA{\textit{Universiy of Milano-Bicocca} \\
Milano, Italy\\
daniela.micucci@unimib.it}
\linebreakand
\IEEEauthorblockN{Maria Teresa Rossi}
\IEEEauthorblockA{\textit{Universiy of Milano-Bicocca} \\
Milano, Italy \\
maria.rossi@unimib.it
}
\and
\IEEEauthorblockN{Riccardo Rubei}
\IEEEauthorblockA{\textit{University of l'Aquila} \\
L'Aquila, Italy\\
riccardo.rubei@univaq.it}
}

\maketitle

\begin{abstract}
IoT systems' complexity and susceptibility to failures pose significant challenges in ensuring their reliable operation. Failures can be internally generated or caused by external factors, impacting both the system's correctness and its surrounding environment.
To investigate these complexities, various modeling approaches have been proposed to raise the level of abstraction, facilitating automation and analysis. Failure-Logic Analysis (FLA) is a technique that helps predict potential failure scenarios by defining how a component's failure logic behaves and spreads throughout the system. However, manually specifying FLA rules can be arduous and error-prone, leading to incomplete or inaccurate specifications.
In this paper, we propose adopting testing methodologies to improve the \emph{completeness} and \emph{correctness} of these rules. How failures may propagate within an IoT system can be observed by systematically injecting failures, while running test cases to collect evidence useful to add, complete and refine FLA rules. 
\end{abstract}


\section{Introduction}
\label{sec:introduction}

IoT systems may experience different failures, either internally generated or caused by the surrounding environment~\cite{KIRCHHOF2022111087}. Such failures may affect not only the correctness of the system, but also the environment in which it operates. Consider, for instance, Smart Irrigation Systems: they monitor parameters related to weather and soil to irrigate crop fields based on the data collected automatically. A failure affecting the behaviour of those IoT systems may cause a waste of water or loss to the farm's production. Since IoT systems are composed of components of different natures (e.g., temperature/humidity sensors, cloud servers, and irrigation units), studying how failures (e.g., caused by a malfunctioning component) may propagate within a system and impact its behaviour can be highly challenging, further than being of high importance \cite{thesisFelicien,challengesIOTFLA}.

Developing IoT systems is complex due to several reasons. The integration of diverse components, the need to handle real-time data, and the distributed nature of IoT systems are just a few factors contributing to this complexity. Several modeling approaches have been proposed over the last few years to raise the level of abstraction (e.g., \cite{MDE4IoT, CAPS, KIRCHHOF2022111087, MonitorIoT}), promoting the adoption of models for increasing automation and easing analysis. These models help understanding systems behaviour, performance, and potential failure scenarios~\cite{ihirwe2021towards}.

Failure-Logic Analysis (FLA)~\cite{Gallina} is one of the analyses that can be applied to IoT systems. By using FLA, it is possible to define how a component's failure logic shall behave, which can help analyze how failures could potentially spread throughout a system and predict any potential issue. For FLA to work correctly, it is important to have accurate information about how failures may occur within each  component and propagate between components.
FLA relies on the manual specification of rules, that rigorously indicate the different kinds of failures that might occur and how they can propagate throughout the components. Specifying such rules is a strenuous and error-prone process, as identifying all possible fault scenarios and formulating accurate rules is challenging, possibly leading to incomplete or incorrect specifications.

In this paper, we propose to adopt testing methodologies to mitigate the issues related to the \emph{completeness} and \emph{correctness} of the manually specified FLA rules. By systematically introducing failures into an IoT system and running test cases, it is possible to observe how failures propagate. The collected evidence is then used to add, refine, and eliminate FLA rules, better capturing the behavior of the system in failure scenarios.

The paper is organized as follows: In Section \ref{sec:motivatingExample} we provide motivation for this work and present an explanatory example. We describe our approach in Section~\ref{sec:proprosedApproach}. Section~\ref{sec:evaluation} reports a preliminary evaluation of the proposed approach. In Section~\ref{sec:relatedWork}, we discuss related work. Finally, Section~\ref{sec:conclusion} concludes the paper and outlines future work.

\section{Motivation and background}
\label{sec:motivatingExample}


Figure \ref{fig:motivating} represents a Smart Irrigation System (SIS) that includes all the building blocks of a typical \iot system, \ie actuators, monitors, and sensors. The  system analyzes the environmental conditions to automatically irrigate the soil using the classical MAPE-K control loop \cite{1160055,Brun2009}. In particular, each node (represented by the dashed line in Figure \ref{fig:motivating}) is composed of different types of sensors, \ie \textit{Moisture}, \textit{Temperature}, and \textit{Humidity}.

\begin{figure}
    \centering
    \includegraphics[width=0.9\linewidth]{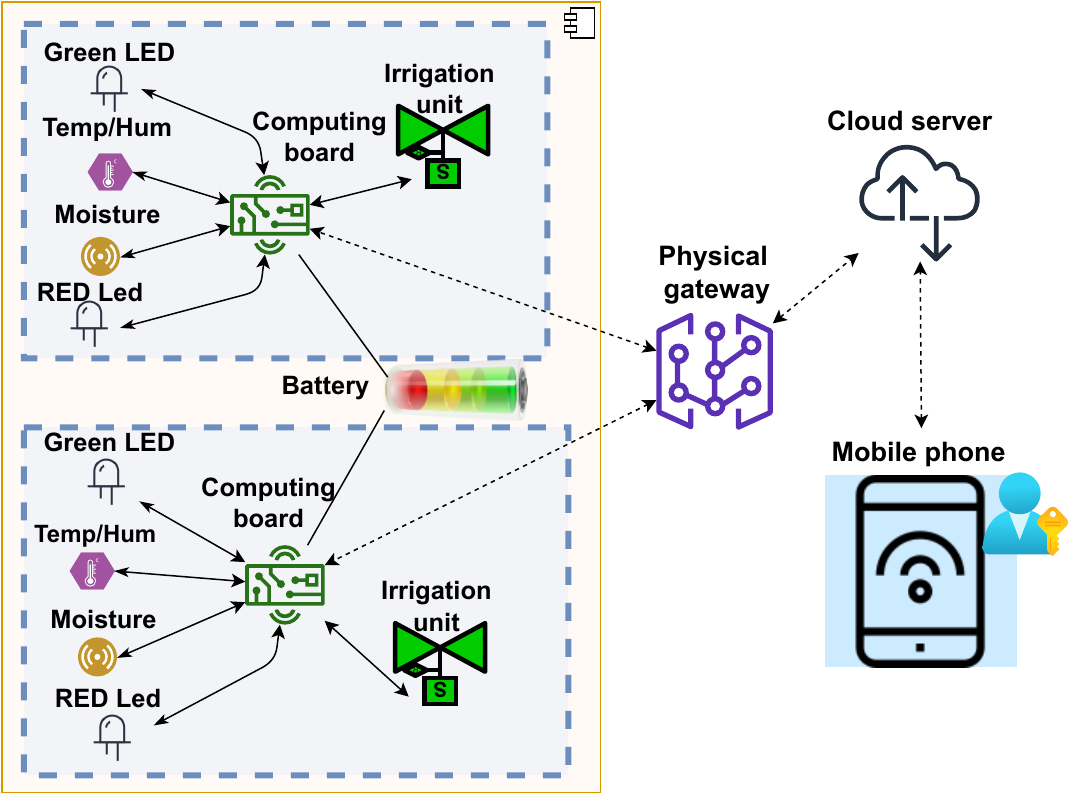}
    
    \caption{The Smart Irrigation System use case.}
    \label{fig:motivating}
\end{figure}

Such sensors collect data at a given node and continuously feed it to the \textit{Computing board}. Based on sensor data, the board decides whether to send a signal straight to the \textit{Irrigation unit} actuator to start or stop the watering process. When the irrigation phase is ended, the \textit{LED} indicators switch from green to red. 
The \textit{Physical gateway} connects each irrigation node to the \textit{Cloud server}, allowing users to remotely control, via \textit{Mobile phone}, the irrigation nodes and analyze sensor data.
Even though the presented system is simple, it represents a real-world application composed of miscellaneous \iot components that can be prone to critical malfunctioning. For instance, the Moisture sensor can send the wrong value, thus causing a waste of water or loss to the farm's production. Similarly, the user can erroneously decide to irrigate the field if the LED is malfunctioning. 
Therefore, failure propagation analysis plays an important role in understanding the system's behavior when it suffers from those faults.  

An early-safety analysis approach has been proposed  by Ihirwe\cite{thesisFelicien},  by relying on Failure-Logic Analysis (FLA)~\cite{Xing2008} mechanisms. 
FLA allows modelers to specify a component's failure logic behavior to help analyze how failures propagate within a system to anticipate possible misbehaviors. 
To be effective, FLA requires accurate knowledge about how failure may behave within the individual components. This can either be by means of propagation or transformation across components. 


Figure \ref{fig:workflow} depicts the failure analysis workflow underpinning the approach proposed by Ihirwe \cite{thesisFelicien}. 
First, the system is modeled by identifying all the needed components and communication channels. Afterward, the user has to check the system's safety by performing a proper failure propagation analysis. 
It is worth noting that the two phases are typically conducted manually with no or limited degree of automation \cite{8411738,6721820}.

\begin{figure}
    \centering
    \includegraphics[width=0.6\linewidth]{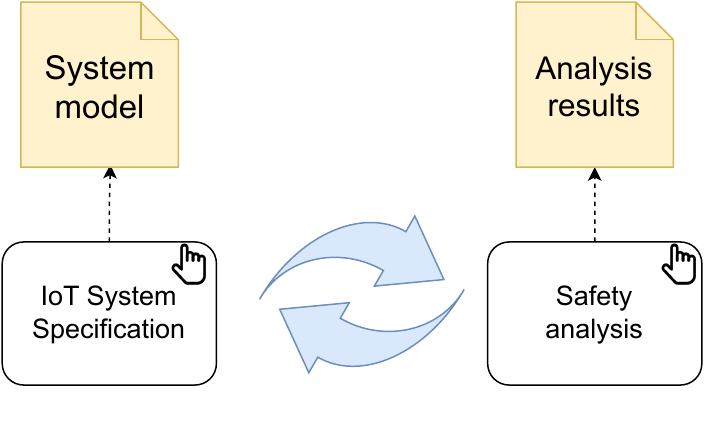}
    \vspace{-4mm}
    \caption{Traditional failure analysis workflow.}
    \label{fig:workflow}
\end{figure}

However, detecting those faults is a daunting task since thoroughly exercising an \iot system requires considering both the system and its environment. Therefore, a task of paramount importance is to detect how failures may propagate using early-safety analysis strategies. Even though several frameworks and techniques are in place~\cite{cristea_building_2022,bures_patriot_2021,215955}, there is a need to verify the correctness of such rules at design time. Thus, the main challenges that need to be addressed when modeling IoT systems while supporting early-safety analysis are as follows:

\begin{itemize}[leftmargin=*]
    \item \textbf{CH1: Detecting fault propagation in \iot systems}
    While fault analysis has been studied in generic software systems \cite{Xing2008}, detecting failures in \iot systems has to consider real-time data that may introduce variability in the conducted analysis. Furthermore, failures that occur at the circuits-level should be considered in the analysis as they cause bugs that impact the source code \cite{10.1145/2858036.2858533,9402092};
    
    \item \textbf{CH2: Verifying the completeness and correctness of fault propagation rules:} Even though fault propagation rules can be specified at the design time, their completeness and correctness cannot be granted \textit{a priori}. 
\end{itemize}







\section{Proposed Approach}
\label{sec:proprosedApproach}

To detect fault propagation (CH1) and verify FLA rules (CH2), we propose a \emph{Model-Based Test-Driven Safety Analysis} approach that allows engineers to identify potential failures and their propagation across components.
%
\begin{figure}[t]
    \centering
    \includegraphics[width=.3\textwidth]{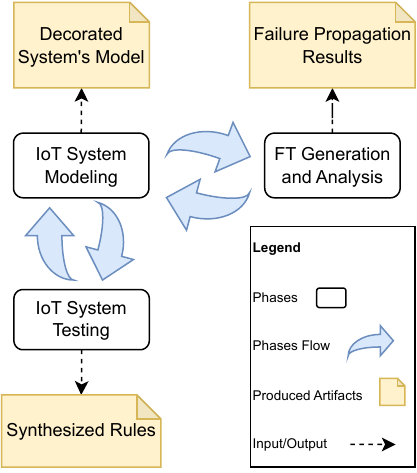}
    \vspace{-1mm}
    \caption{Model-Based Test-Driven Safety Analysis.}
    \label{fig:approach}
\end{figure}
The proposed approach implements and extends the one shown in Figure \ref{fig:workflow} and consists of the three main phases shown in Figure~\ref{fig:approach}. The \emph{IoT System Modeling} phase proposes tool-supported modeling of the IoT system-level architecture and the modeling of the system failure logic behavior. The \emph{Fault-Tree Generation and Analysis} supports the analysis of failure propagation, with reference to the available rules. 
Finally, the \emph{IoT System Testing} phase exploits the information in the model to execute the individual components while injecting failures on input ports and checking how they propagate to output ports. This phase can confirm or disprove the defined failure logic rules (correctness check) and discover new ones (completeness check). 
The outcome of both analysis and testing can be used to refine the system, and its model, to finally achieve a more reliable IoT system. 

In the following, the three phases of the process shown in Figure~\ref{fig:approach} are described in details.

\subsection{IoT System Modeling} \label{sec:SM} In this phase, modelers specify the architecture of the IoT system and the failure-logic behaviour, as detailed below.

\subsubsection{IoT system-level architecture}

The proposed modeling approach runs on top of CHESSIoT \cite{thesisFelicien}, a model-driven environment to support the design and analysis of IoT systems. CHESSIoT provides a UML/SysML profile extension to reflect the constructs and semantics present in IoT system-level architectures. The CHESSIoT system-level modeling language was designed to satisfy the high-level specifications of a typical IoT system, supporting a multi-layered specification  from the low-level edge layer to the fog layer and the cloud.

The language extends the SysML modeling language in terms of new IoT-specific stereotypes and their interrelations. Ports enable interactions among components and are fundamental for determining error propagation paths.
Figure \ref{fig:CHESSIoTSystemMetamodel} presents the CHESSIoT system-level meta-model. It permits to specify IoT systems as 
a collection of physical devices and entities connected to collect, process, send, receive, and store data. The \textit{IoTElement} represents physical entities, ranging from microcontrollers at the thing layer to cloud servers. The modeling layers can be grouped into \textit{edge}, \textit{fog}, and \textit{cloud}.

\begin{figure}[t!]
	\centering
	\includegraphics[width=\linewidth]{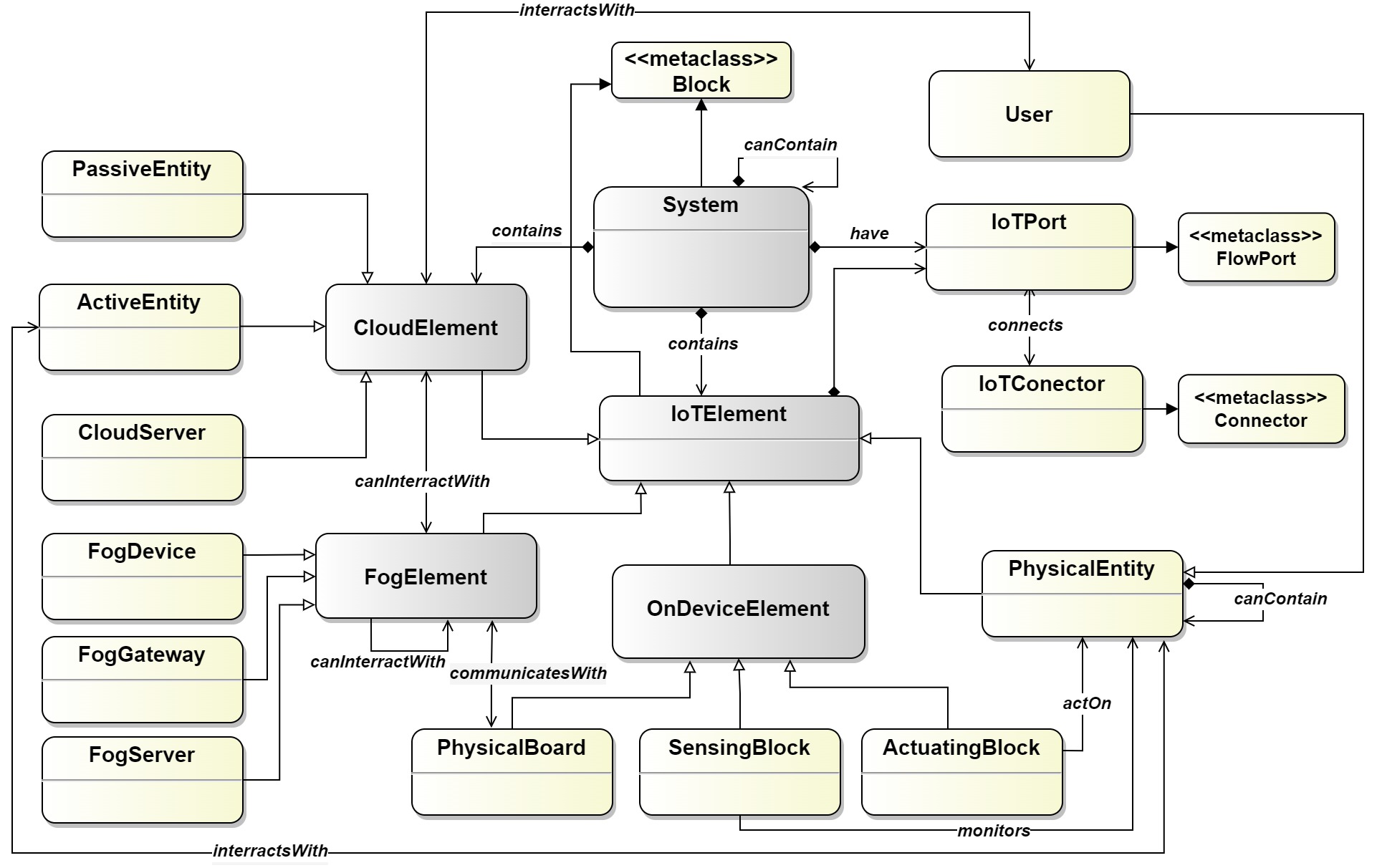}
	\caption{CHESSIoT System-level meta-model \cite{thesisFelicien}.}
	\label{fig:CHESSIoTSystemMetamodel}
\end{figure}

\textit{OnDeviceElements} are low-level IoT devices that contribute to the system's functional behavior, while \textit{PhysicalBoard} represents hardware controllers and \textit{PhysicalEntity} is any physical object or environment. \textit{Fog devices} perform preliminary computations and convey results to on-device elements, with storage and processing capacities varying depending on the use case and hardware and software features.

On the cloud layer, devices operate at the cloud level and contribute to the overall functionality of the system. \textit{Consumer entities} can be \textit{active} or \textit{passive}, with active consumer entities being computer-running software to monitor and control sensors remotely, and passive consumer entities being traffic light actuators.

\subsubsection{Failure-Logic behavior modeling and analysis} \label{sec:flamodeling}

Once the IoT system model is defined, the safety engineer derives and annotates the failure behavior rules for each modeled component by following the Failure Propagation Transformation Calculus (FPTC) \cite{WALLACE200553} notation. Based on its nature, a component can propagate a failure (carrying a failure from input to output), transform a failure (changing the nature of a failure from input to output), act as a source of failure (creating a failure despite no failure in input), or act as a sink (avoiding the failure to be either propagated or transformed). 

The following three abstract categories of failure types are assessed: \textit{service provision failures}, such as the omission or commission of the output; \textit{timing failures}, such as the early or late delivery of the output; and \textit{value domain failures}, such as the output value being out of a valid range, stuck, or exhibiting erratic behavior. In addition, a \textit{noFailure} annotation is used to indicate a no-failure type at the input port.  Table \ref{tab:flatypes} shows different failure types and their descriptions.

\begin{table}[h!]
  \scriptsize
  \centering
\caption{Failure types.}
\vspace{-2mm}
  \begin{tabular}{|c|c|}\hline
    \textbf{\texttt{Failure type}} & \textbf{\texttt{Description}}\\\hline
    \texttt{Early} & \texttt{Output provided too early}\\
    \texttt{Late} &  \texttt{Output provided too late}\\
    \texttt{ValueCoarse} & \texttt{Output out of range}\\
    \texttt{ValueSubtle} & 
    \texttt{Output in-range but erroneous}\\
    \texttt{Omission} & \texttt{Output expected but not provided}\\
    \texttt{Commission} & \texttt{Output provided but not expected}\\\hline
  \end{tabular}

  \label{tab:flatypes}
\end{table}

As previously mentioned, component failures can be propagated or transformed:

\begin{itemize}
    \item \textbf{\textit{Failure propagation:}} It occurs in a component when a single input port failure condition is directly transferred to its output ports without changing its nature. For instance, Equation \ref{eqPropagation} shows a simple example of a failure propagation of $failure_1$ from port $p_{(in)}$ to port $p_{(out)}$ of a simple component. Propagation also occurs between two connected components when a failure condition at the output port of the preceding component is transferred to the input port of the following component. 
    
    \begin{equation} \small \label{eqPropagation}
         p_{(in)}.failure_1 \rightarrow p_{(out)}.failure_1;
    \end{equation}
    
    As an example from our specific scenario, this can happen when a board gets erroneous data from the sensors (i.e., \texttt{ValueCoarse}) and sends it directly to its output ports. This scenario would be expressed as in Equation \ref{ValueCoarseBD}, where $Bd_{(in)}$ and $Bd_{(out)}$ are the input and the output ports of the board, respectively.

    \begin{equation}\label{ValueCoarseBD}
    Bd_{(in)}.valueCoarse \rightarrow Bd_{(out)}.valueCoarse;
    \end{equation}

    
    \item \textbf{\textit{Failure transformation:}} It occurs within a component when a failure condition at the input port is converted into another type before reaching the output port. An example is shown in Equation  \ref{eqtransformation1}. A failure transformation can also occur when more than one failure expression of any type, except a \texttt{NoFailure} or \textit{wildcard} at multiple input ports, is transmitted on a single output port (see Equation \ref{eqtransformation2}). Even if the failure has the same type, the fact that the component converts two failures at its input ports to a single failure at the output port is regarded as a failure transformation.
    
    \begin{equation} \small \label{eqtransformation1}
     p_{(in)}.failure_{(in)} \rightarrow p_{(out)}.failure_{(out)};
    \end{equation}
    \begin{multline} \small \label{eqtransformation2}
        p_{(in_1)}.failure_1,..,p_{(in_N)}.failure_N \rightarrow \\ p_{(out)}.failure_{(out)};
    \end{multline}
\end{itemize}

To make an example of failure propagations and transformations, let us consider the explanatory irrigation system with two motors controlled by a relay driver. The relay driver enables them to turn on and off depending on the location to be irrigated. To control the relay driver, the computer board sends the analog signal through two relay driver-controlling ports. To define the failure behavior of the irrigation unit, we must first understand the variety of failure scenarios that can occur with an irrigation unit. For example, two input ports may not get a signal from the board, causing the relay driver to be unable to switch on and off the motors. Another example is when the signal arrives at the input port later or earlier than anticipated. As a result, the relay will unexpectedly turn on and off the motors. 

Table \ref{tab:IUFLAtypes} shows a sample of the failure rules that specify the number of failure situations for the irrigation unit, including the ones described above. Note, $Irr_{(in_1)}$ and $Irr_{(in_2)}$ are defined as input ports, while $Irr_{(out_1)}$ and $Irr_{(out_2)}$ are defined as output ports.

\begin{table*}[h]
    \centering
    \scriptsize
    \caption{Sample FLA rules of the irrigation unit.}
    \begin{tabular}{|p{40em}|p{29em}|}\hline
        \scriptsize{\textbf{Rule}} & \scriptsize{\textbf{Description}} \\\hline\hline
        $Irr_{(in_1)}.omission, Irr_{(in_2)}.omission \rightarrow Irr_{(out_1)}.omission, Irr_{(out_2)}.omission$
        & The input ports receive no signal, causing the relay driver to be unable to turn on/off the motors. \\\hdashline[1pt/1pt]
        $Irr_{(in_1)}.early, Irr_{(in_2)}.early \rightarrow Irr_{(out_1)}.commission, Irr_{(out_2)}.commission
        $
        & The input ports receive the signal earlier than expected, causing the relay driver to unexpectedly turn on/off the motors.\\\hdashline[1pt/1pt]
        $Irr_{(in_1)}.late, Irr_{(in_2)}.late \rightarrow Irr_{(out_1)}.commission, Irr_{(out_2)}.commission$& The input ports receive the signal later than expected, causing the relay driver to unexpectedly turn on/off the motors.\\\hdashline[1pt/1pt]
        $Irr_{(in_1)}.valueSubtle, Irr_{(in_2)}.valueSubtle \rightarrow Irr_{(out_1)}.early, Irr_{(out_2)}.early$& The input ports receive an erroneous but in-range signal, causing the relay driver to turn on/off the motors earlier than expected.\\\hdashline[1pt/1pt]
        $Irr_{(in_1)}.valueSubtle, Irr_{(in_2)}.valueSubtle \rightarrow Irr_{(out_1)}.late, Irr_{(out_2)}.late$& The input ports receive an erroneous but in-range signal, causing the relay driver to turn on/off the motors later than expected.\\\hline
    \end{tabular}
    \label{tab:IUFLAtypes}
\end{table*}

When modeling the IoT system's Failure-Logic behavior is finished, the FLA analysis can be executed. This analysis considers the annotated CHESSIoT model and transforms it into an FLA model \cite{Gallina}. The transformation calculates a complete system's failure behavior starting from the failure behavior rules of the system's composite components and their interconnections. This, in turn, means that the failure behaviors of composite elements are also determined by the failure behaviors of their individual simple components.

\begin{figure}[t!]
  \centering
  \includegraphics[width=\linewidth]{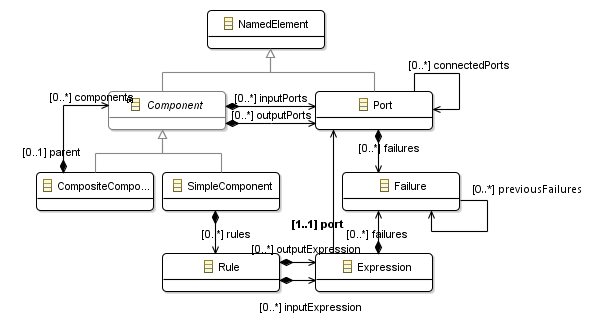}
  \vspace{-7mm}
    \caption{FLA meta-model \cite{thesisFelicien}.}
    \label{fig:flamm}
\end{figure}

As shown in the FLA meta-model in Figure \ref{fig:flamm},  FLA models consist of \textit{composite components}, representing sub-systems containing one or more sub-components. These components do not possess failure behavior by themselves; instead, they rely on their sub-components to determine their failure conditions. On the other hand, a simple component represents a functional component whose failure may contribute to a system failure. Each component contains input and output ports with their corresponding failure rules.

\subsection{Fault-Tree Generation and Analysis} \label{sec:fta}

The Fault-Tree Analysis (FTA) \cite{Xing2008} aims to graphically analyze the system's final failure behavior based on the FLA input. Fault trees depict the system failure logic outcomes in a tree structure, making it simple to navigate and trace influences from a system-level danger to specific failures from system components and sub-components. In addition to that, it is also possible to perform analyses on it to determine minimal failure events that are required to trigger such hazards 



The Fault-Tree generation is performed through a series of model-to-model transformations from the FLA model to a series of Fault-Tree (FT) models. 
An FT is generated for each of the failures that propagate to the targeted output port of the system, and contains logical networks of events and corresponding gates that together form a failure representation tree, reflecting the system's failure behavior set by the user and the system's functional architecture. Each FT event has its own unique identity in the tree and can be of type basic, intermediate, external, or undeveloped, depending on the stage at which it manifests.

In the FT generation process, each FT is built recursively. A top event is initially generated due to the failure's propagation to the system output port. In terms of logical gates used in the FT, only \textbf{AND} and \textbf{OR} gates are adopted. The events gate is created systematically.  An \textbf{AND} gate is used to indicate a failure transformation from an input to an output port of a component. On the other hand, an \textbf{OR} gate is used to depict a failure propagation case. The \textbf{OR} gate can also depict a scenario in which one or more failure outputs from distinct components are passed to the input of the following component.

The intermediate events are created and populated into the FT based on the failure expressions and the components they are assigned to. The FT population involves a recursive transformation process in which components, ports, and their corresponding rules are recursively parsed. So, at this stage, the only crucial stopping case is reached when the transformation hits a condition matching a basic failure, an underdeveloped failure (i.e., an insufficient source failure), or an externally injected failure.
\begin{figure}[b]
  \centering
  \includegraphics[width=.9\linewidth]{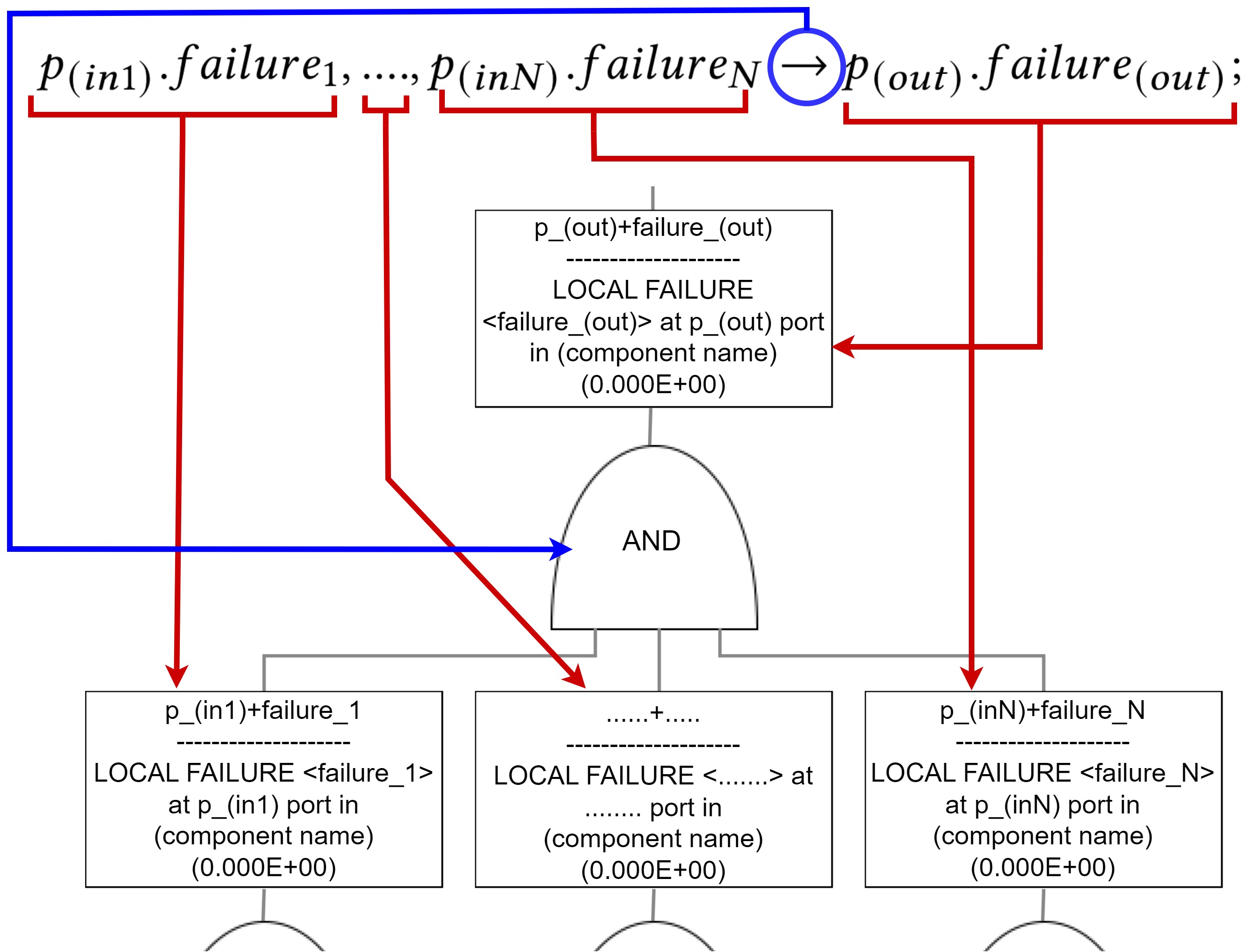}
    \caption{FT corresponding to Equation \ref{eqtransformation2}.}
    \label{fig:eqtransformation2FT }
\end{figure}
\begin{figure*}
    \centering
    \includegraphics[width=.8\textwidth]{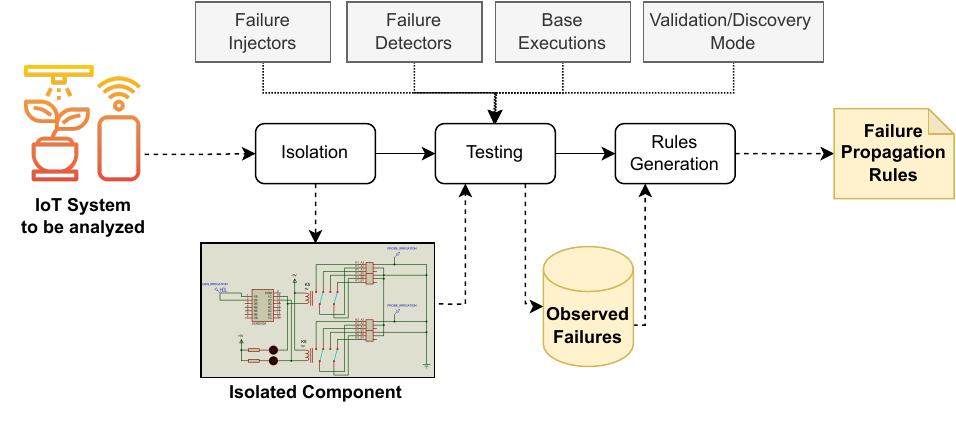}
    \vspace{-5mm}
    \caption{IoT System's testing process.}
    \label{fig:testing_process}
\end{figure*}
For instance,  Figure \ref{fig:eqtransformation2FT } depicts a simple transformation example with indications showing a transformation mapping of Equation \ref{eqtransformation2}. From the example, each of the output expressions is mapped to an output event of a logical combination of the input expressions. Each input expression is mapped to an event, and the type of such event is determined by the expression condition. In addition to that, the logical gates are defined based on the nature of the input expressions to satisfy the failure propagation and transformation concepts.

As the system gets bigger and more complex, which in turn requires a large number of rules to better cover all possible failure scenarios, the generated FTs inevitably become even bigger and harder to grasp. To tackle such a challenge, different analysis mechanisms are used to systematically extract meaningful insight from the generated tree. CHESSIoT supports \textit{``Qualitative"} fault tree analysis mechanisms in which only the essential FT representations are kept. This process involves the removal of internal component failure propagations, external component-to-component failure propagations, and basic event redundancies. In addition, CHESSIoT also supports \textit{``Quantitative"} analysis that automatically calculates the failure probabilities of an entire system from its constituent parts’ failure probabilities.

\subsection{IoT System Testing}\label{sec:fpr}

The third phase of the approach shown in Figure~\ref{fig:approach} concerns testing the modeled IoT system to confirm or disprove the defined failure logic rules (correctness check) and discover
new ones (completeness check). This phase is guided by the information collected from the system's model and consists of three main activities: the \emph{Isolation} of the components to be exercised, the \emph{Testing} of the isolated components to collect observations about how failures are propagated from inputs to outputs, and the \emph{Rules Generation} from the observations, as shown in Figure \ref{fig:testing_process}.

\begin{figure}[h!]
    \centering
    \includegraphics[width=.48\textwidth]{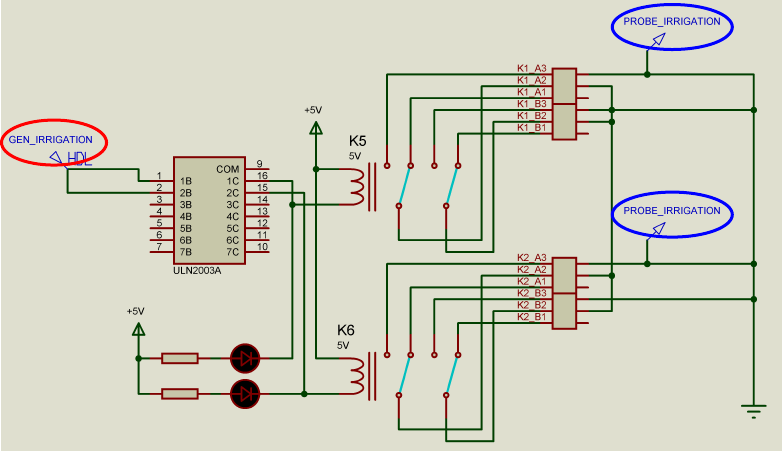}
    \caption{Irrigation unit component isolated.}
    \label{fig:isolated}
\end{figure}

The \emph{Isolation} activity isolates the component under test from the rest of the system using stubs and probes. The stubs are connected to the input ports of the component while removing the original connections. The probes are connected to the output ports of the components, again removing the original connections. In this configuration, the stubs generate the input values in a controlled and coordinated way, whereas monitoring probes capture and record the output values produced by the component under test. Figure~\ref{fig:isolated} shows how the explanatory component \emph{Irrigation unit} is isolated to support the discovery of failure propagation patterns. Isolation is a simple activity performed visually on the model.

The \emph{Testing} activity consists of exercising the isolated component through the stubs to log evidence about how failures are propagated through the probes. Since the inputs of an isolated component are fully controllable and its outputs can be fully observed, it is possible to systematically generate tests that include failures in the inputs and observe how and if they propagate to the outputs. 

The strategy to observe how failures propagate is based on a variant of \emph{differential testing} \cite{Barany2018,Chen2023}. In differential testing, the same inputs are executed on two comparable implementations (e.g., two compilers for the same language), and the outputs are directly compared to discover possible faults. In this case, we start from a \emph{base execution} \texttt{t=(I,O)}, where \texttt{I} is a set of time series values, each one representing a sequence of input values for an input port, and \texttt{O} is a set of time series values, each one representing a sequence of output values observed for an output port. Figure \ref{fig:input-output} shows an example of a base execution of the Irrigation Unit component. The time series provided by the stub in input activates the component from second 15.00001 to second 30.00001, by sending 5 Volts to the circuits to turn on the water fans and the LED associated with that input port; instead, when the value is set to 0, the component is not active as not stimulated by any Volt. The time series in output reflects the behavior instrumented in input, as the component results active, for each output port, from second 15.00004 to second 30.00007 (differences with respect to input are minimal and depend on the precision of Proteus tool\footnote{\url{https://www.labcenter.com/iotbuilder/}}, which is used for this work as design and simulation environment). 

\begin{figure}
    \centering
    \includegraphics[trim=0.0cm 0.0cm 0.0cm 0.0cm, clip=true, width=.47\textwidth]{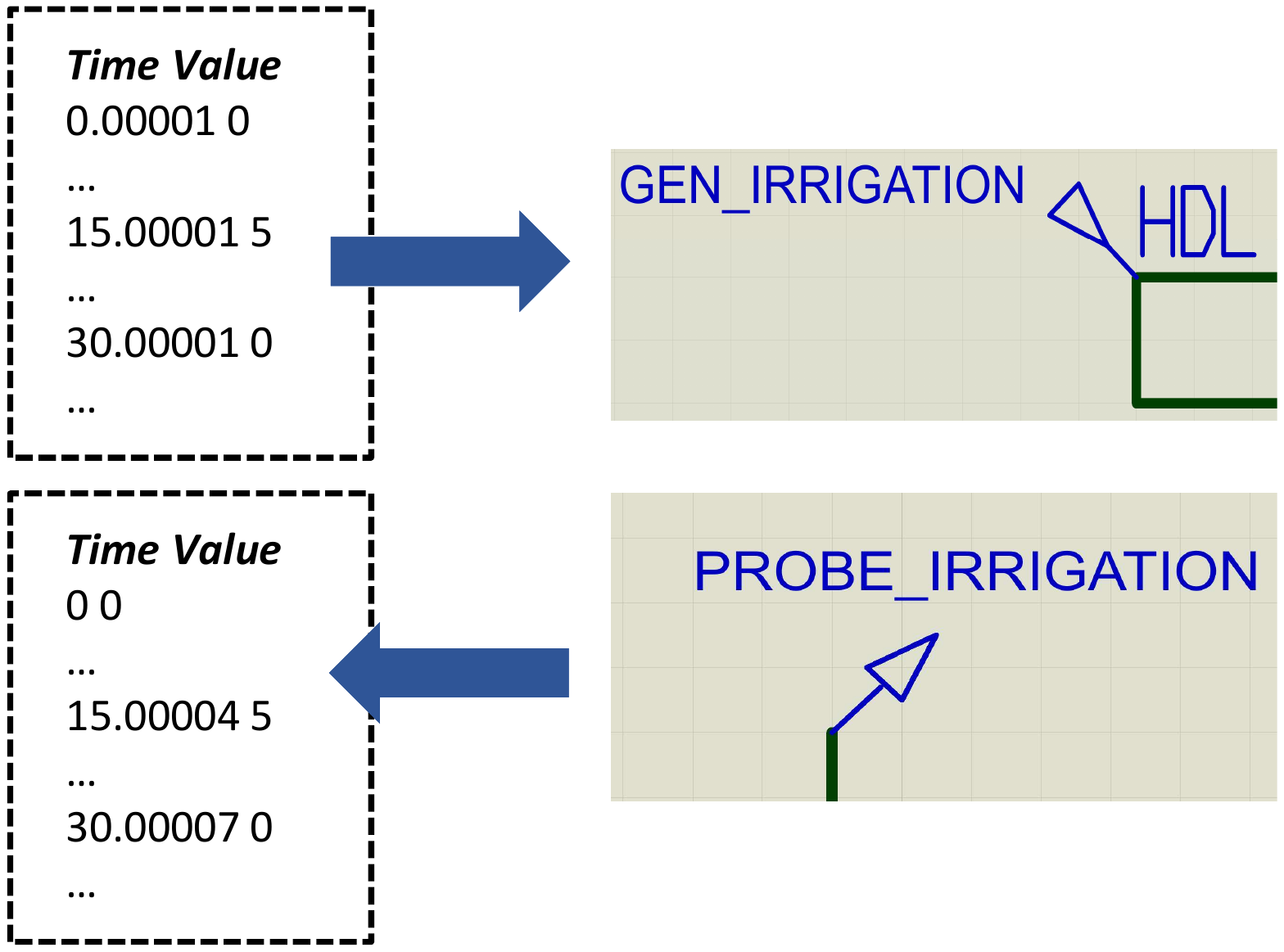}
    \vspace{-5mm}
    \caption{Input \& output time series of Irrigation unit component.}
    \label{fig:input-output}
\end{figure}

To discover how failures propagate, our approach automatically modifies the inputs $I$ used in a base execution $t$ by systematically injecting failures of the given types on the inputs. The approach then executes the modified inputs $I'$ and collects the outputs, namely $O'$. The comparison of the output produced by the base execution $O$ and the output generated by the mutated execution $O'$ reveals if and how the input failure(s) propagated to the output. 

We need two main elements for each supported failure type to execute this process: a failure injector and a failure detector. The \emph{failure injector} is a function that, given an input time series, modifies its values to obtain a minimally modified time series that includes the failure of the given type. The \emph{failure detector} is a function that, given two time series, one obtained from the base execution and another obtained from a mutated execution, can tell if the failure of the given type is present in the output. 

Table \ref{tab:failures_inj_det} summarizes the failure injectors and the failure detectors defined for the fault types currently supported by our implementation for IoT systems. 
Note that both injectors and detectors can be parameterized with respect to the actual use case, to reflect the characteristics and volatility of the signals. $TS$ and $TS^{'}$ represent base and mutated time series, respectively, $\varepsilon_{t}$ and $\varepsilon_{v}$ determine the tolerance on timing and values variations, and $[v_{min}, v_{max}]$ are the range of accepted values for the considered use case.  As an example, an \texttt{Early} failure injector may mutate the base execution of Figure \ref{fig:input-output}, by changing the component activation time from second 15.00001 to 12.00001, whereas an \texttt{Early} detector may compare base and mutated executions to determine how this propagates in output.

\begin{table*}[h]
    \centering
    \scriptsize
    \caption{Failure Injectors \& Detectors.}
    \begin{tabular}{|p{6em}|p{26em}|p{37em}|}\hline
		\scriptsize{\textbf{Type}} & \scriptsize{\textbf{Injectors}} & \scriptsize{\textbf{Detectors}}\\\hline\hline
        \texttt{Early} &  $(t_i, v_i) \in TS \rightarrow (t_i - x, v_i) \in TS^{'}$, where $(x > \varepsilon_{t})$
				& $(t_i, v_i) \in TS, (t_j, v_j) \in TS^{'} \rightarrow 
    true$, if $(v_i = v_j) \land (t_j < t_i - \varepsilon_{t})$ \\\hline
    
				\texttt{Late} & $(t_i, v_i) \in TS \rightarrow (t_i + x, v_i) \in TS^{'}$, where $(x > \varepsilon_{t})$  
				& $(t_i, v_i) \in TS, (t_j, v_j) \in TS^{'} \rightarrow 
    true$, if $(v_i = v_j) \land (t_j > t_i + \varepsilon_{t})$ \\\hline
    
			  \texttt{ValueCoarse} & $(t_i, v_i) \in TS \rightarrow  (t_i, v_i \pm x) \in TS^{'}$, where $ (x > \varepsilon_{v}) \land ((v_i - x <v_{min}) \lor (v_i + x >v_{max}))$ 
				& $(t_i, v_i) \in TS, (t_j, v_j) \in TS^{'} \rightarrow 
    true$, if $(t_i = t_j) \land ((v_j < v_{min}) \lor (v_j > v_{max}))$ \\\hline

				\texttt{ValueSubtle} & $(t_i, v_i) \in TS \rightarrow (t_i, v_i \pm x) \in TS^{'}$, where 
                $ (x > \varepsilon_{v}) \land (v_{min} \leq \lvert v_i \pm x \rvert \leq v_{max})$ 
				& $(t_i, v_i) \in TS, (t_j, v_j) \in TS^{'} \rightarrow 
    true$, if $(t_i = t_j) \land (\lvert v_i - v_j \rvert > \varepsilon_{v}) \land (v_{min} \leq  v_j \leq v_{max})$ \\\hline 
    \end{tabular}
    \label{tab:failures_inj_det}
    \vspace{-2mm}
\end{table*}

These elements are combined into a fully automated testing process that repeatedly instantiates the isolated component, generates the mutated execution (using the failure injectors), runs the component with the mutated execution, collects observations checking for the presence of failures (using the failure detectors), and destroys the instance of the isolated component. This process is repeated multiple times for all the combinations of failures to be investigated.   


Additional data can be collected by considering multiple base executions. In particular, we envision the possibility to start from a set of base executions that cover the various possible states of the components under test, so that the propagation of failures can be studied in multiple contests (i.e., states). For the moment, we assume the user of the approach shall define the states that must be used in this process, and thus provide a set of base executions that cover the relevant states. In the future, we would like to explore the automatic generation of the base executions.

At the end of this process, for each combination of input failures $p_{(in1)}.failure^{i_1}_{(in1)},..,p_{(inN)}.failure^{i_N}_{(inN)}$, a set of observations $Obs_i$ where $Obs_i \in \{failure^1 \ldots failure^k\} \cup NoFailure$, are available. 

This testing phase can be configured to work in two different modes: (i) \textit{validation} of existing failure propagation rules (validation mode) or (ii) \textit{discovery} of new failure propagation rules (discovery mode).

The \emph{validation mode} provides a cost-effective targeted exploration of the combinations of failures, and their propagation. In particular, for each failure propagation rule $p_{(in1)}.failure_{(in1)},..,p_{(inN)}.failure_{(inN)} \rightarrow p_{(out)}.failure_{(out)}$, the validation mode investigates how failures propagate to the output when the failures in the input ports are consistent with the left-hand side of the rule, i.e., $failure_{(in1)},..,failure_{(inN)}$. The final set of derived rules will either confirm or disprove the existing rules.

The \emph{discovery mode} is a more expensive but systematic exploration of the possibile combinations of failures to derive rules about their propagation. Given a set of $k$ failure types $failure^1 \ldots failure^k$ and $N$ input ports, this mode investigates how, and if, failures propagate for every combination of input failures $p_{(in1)}.failure^{i_1}_{(in1)},..,p_{(inN)}.failure^{i_N}_{(inN)}$, where each tuple indicates the failures present in the input ports, and $failure^i$ is any of the considered failure types or \texttt{NoFailure}. For instance, if the \texttt{Late (L)}, and \texttt{Early (E)} failure types are investigated for two input ports, six combinations of inputs failures ($\langle L, - \rangle, \langle -, L\rangle, \langle R, -\rangle, \langle -, R\rangle, \langle L, R\rangle, \langle R, L\rangle $, where $-$ represents \texttt{NoFailure}),  obtained by every possible permutation of the considered failures, are considered. The final set of rules will provide a comprehensive view about how failures are propagated by the considered component.

In practice, the validation mode constraints the discovery to the subset of failures used in the rules to be validated, while the discovery mode considers every possible combination of failures. 

Finally, the \emph{Rules Generation} activity of Figure \ref{fig:testing_process} consists of the generation of the actual set of failure propagation rules. This is done by extracting the set of failures types $f_1, \ldots f_p$ present in a set of observations $Obs_i$ associated with a same pattern of failures $p_{(in1)}.failure^{i_1}_{(in1)},..,p_{(inN)}.failure^{i_N}_{(inN)}$ and generating the rule $p_{(in1)}.failure^{i_1}_{(in1)},..,p_{(inN)}.failure^{i_N}_{(inN)} \rightarrow f_1 or \ldots or f_p$, which is finally encoded in the failure analysis tool as a set of $p$ non-deterministc rules $p_{(in1)}.failure^{i_1}_{(in1)},..,p_{(inN)}.failure^{i_N}_{(inN)} \rightarrow f_i$, where $i=1\ldots p$.  


The discovered rules may confirm or disprove the existing failure propagation rules and discover new ones. This phase may thus trigger an evolution of the system's model by refining the system failure logic behaviour definition.
From this evolution, the whole process may be re-triggered to achieve satisfactory reliability for the candidate IoT system.



\section{Evaluation}
\label{sec:evaluation}





To evaluate our approach, we selected the Irrigation unit component, introduced in Figure \ref{fig:testing_process} and shown in more details in Figure~\ref{fig:isolated}. 
The component, designed with Proteus, comprises two main circuit streams that decide the activation of two water fans and two LEDs (each circuit controls a water fan and a LED). Once isolated, the input signals of the circuit are generated by the \textit{GEN\_IRRIGATION} stub (red circle in Figure~\ref{fig:isolated}), while the output signals are linked to the \textit{PROBE\_IRRIGATION} probes that record the data produced by the circuits (blue circles in Figure~\ref{fig:isolated}).

We use the base execution described in Figure~\ref{fig:input-output} as a basis to study failure propagation. In this scenario, \textit{both the water fans and LEDs start as turned off, then they are turned on for a fixed timespan by an external request, and finally they are turned off again}. 
As mutated scenarios, we considered every possible combination of the currently supported failure types (i.e., \texttt{Early}, \texttt{Late}, \texttt{ValueCoarse}, and \texttt{ValueSubtle}, from Table \ref{tab:flatypes}) for the two input ports available in the component. Every combination was repeated 3 times, in order to capture variations in the executions, for a total of 48 different mutated scenarios produced by our \textit{injectors}.
We injected each combination separately, ran the mutated scenario, and compared the results via \textit{detectors} to discover how failures propagate. 


The experiment's results are summarized in Table \ref{tab:test-results}. The column labeled \emph{IN Failures} shows the combinations of failures that were injected into the two input ports, while the column labeled \emph{OUT Failures} displays the combination of failures that were observed on the two output ports. As each combination of input failures was executed multiple times, it was possible to observe various combinations of failures on the output. For example, when the first and second unit ports were given the failures \texttt{Early} and \texttt{ValueSubtle}, respectively, two possible combinations of output failures were identified: \texttt{Early-Late} and \texttt{Early-NoFailure}.

The \textcolor{OliveGreen}{\texttt{green color}} highlights a failure type that propagates unchanged from the input to the output. The \textcolor{Red}{\texttt{red color}} indicates a failure type that is transformed as a failure of a different type. Finally, the \textcolor{Blue}{\texttt{blue color}} determines failures that are masked by the implementation and thus do not propagate to the output.

\begin{table*}[]
\centering
\caption{Failures Propagation in Irrigation unit experiment.}
\begin{tabular}{c|c}
\hline
\textbf{IN Failures}            &  \textbf{OUT Failures}                                         \\ \hline
\multicolumn{1}{l|}{\texttt{Early - Early}}             &  \multicolumn{1}{l}{\textcolor{OliveGreen}{\texttt{Early}} \texttt{-} \textcolor{OliveGreen}{\texttt{Early}}}                     \\ 

\multicolumn{1}{l|}{\texttt{Early - Late}}              & \multicolumn{1}{l}{\textcolor{OliveGreen}{\texttt{Early}} \texttt{-} \textcolor{OliveGreen}{\texttt{Late}}}                      \\ 

\multicolumn{1}{l|}{\texttt{Early - ValueCoarse}}       & \multicolumn{1}{l}{\textcolor{OliveGreen}{\texttt{Early}} \texttt{-} \textcolor{Red}{\texttt{Late}}, \textcolor{OliveGreen}{\texttt{Early}} \texttt{-} \textcolor{Blue}{\texttt{NoFailure}}}                      \\ 

\multicolumn{1}{l|}{\texttt{Early - ValueSubtle}}       & \multicolumn{1}{l}{\textcolor{OliveGreen}{\texttt{Early}} \texttt{-} \textcolor{Red}{\texttt{Late}}, \textcolor{OliveGreen}{\texttt{Early}} \texttt{-} \textcolor{Blue}{\texttt{NoFailure}}}                      \\ 

\multicolumn{1}{l|}{\texttt{Late - Early}}              & \multicolumn{1}{l}{\textcolor{OliveGreen}{\texttt{Late}} \texttt{-} \textcolor{OliveGreen}{\texttt{Early}}}                      \\ 

\multicolumn{1}{l|}{\texttt{Late - Late}}               & \multicolumn{1}{l}{\textcolor{OliveGreen}{\texttt{Late}} \texttt{-} \textcolor{OliveGreen}{\texttt{Late}}}                      \\ 

\multicolumn{1}{l|}{\texttt{Late - ValueCoarse}}        & \multicolumn{1}{l}{\textcolor{OliveGreen}{\texttt{Late}} \texttt{-} \textcolor{Red}{\texttt{Late}}, \textcolor{OliveGreen}{\texttt{Late}} \texttt{-} \textcolor{Blue}{\texttt{NoFailure}}}                      \\ 

\multicolumn{1}{l|}{\texttt{Late - ValueSubtle}}        & \multicolumn{1}{l}{\textcolor{OliveGreen}{\texttt{Late}} \texttt{-} \textcolor{Red}{\texttt{Late}}, \textcolor{OliveGreen}{\texttt{Late}} \texttt{-} \textcolor{Blue}{\texttt{NoFailure}}}                      \\ 

\multicolumn{1}{l|}{\texttt{ValueCoarse - Early}}       & \multicolumn{1}{l}{\textcolor{Red}{\texttt{Late}} \texttt{-} \textcolor{OliveGreen}{\texttt{Early}}, \textcolor{Blue}{\texttt{NoFailure}} \texttt{-} \textcolor{OliveGreen}{\texttt{Early}}}                      \\ 

\multicolumn{1}{l|}{\texttt{ValueCoarse - Late}}        & \multicolumn{1}{l}{\textcolor{Red}{\texttt{Late}} \texttt{-} \textcolor{OliveGreen}{\texttt{Late}}, \textcolor{Blue}{\texttt{NoFailure}} \texttt{-} \textcolor{OliveGreen}{\texttt{Late}}}                      \\ 

\multicolumn{1}{l|}{\texttt{ValueCoarse - ValueCoarse}} & \multicolumn{1}{l}{\textcolor{Blue}{\texttt{NoFailure}} \texttt{-} \textcolor{Red}{\texttt{Late}}, \textcolor{Red}{\texttt{Late}} \texttt{-} \textcolor{Blue}{\texttt{NoFailure}}, \textcolor{Red}{\texttt{Late}} \texttt{-} \textcolor{Red}{\texttt{Late}}, \textcolor{Blue}{\texttt{NoFailure}} \texttt{-} \textcolor{Blue}{\texttt{NoFailure}}}                      \\ 

\multicolumn{1}{l|}{\texttt{ValueCoarse - ValueSubtle}} & \multicolumn{1}{l}{\textcolor{Blue}{\texttt{NoFailure}} \texttt{-} \textcolor{Red}{\texttt{Late}}, \textcolor{Red}{\texttt{Late}} \texttt{-} \textcolor{Blue}{\texttt{NoFailure}}, \textcolor{Red}{\texttt{Late}} \texttt{-} \textcolor{Red}{\texttt{Late}}, \textcolor{Blue}{\texttt{NoFailure}} \texttt{-} \textcolor{Blue}{\texttt{NoFailure}}}                      \\ 

\multicolumn{1}{l|}{\texttt{ValueSubtle - Early}}       & \multicolumn{1}{l}{\textcolor{Red}{\texttt{Late}} \texttt{-} \textcolor{OliveGreen}{\texttt{Early}}, \textcolor{Blue}{\texttt{NoFailure}} \texttt{-} \textcolor{OliveGreen}{\texttt{Early}}}                      \\ 

\multicolumn{1}{l|}{\texttt{ValueSubtle - Late}}        & \multicolumn{1}{l}{\textcolor{Red}{\texttt{Late}} \texttt{-} \textcolor{OliveGreen}{\texttt{Late}}, \textcolor{Blue}{\texttt{NoFailure}} \texttt{-} \textcolor{OliveGreen}{\texttt{Late}}}                      \\ 

\multicolumn{1}{l|}{\texttt{ValueSubtle - ValueCoarse}} & \multicolumn{1}{l}{\textcolor{Blue}{\texttt{NoFailure}} \texttt{-} \textcolor{Red}{\texttt{Late}}, \textcolor{Blue}{\texttt{NoFailure}} \texttt{-} \textcolor{Blue}{\texttt{NoFailure}}, \textcolor{Red}{\texttt{Late}} \texttt{-} \textcolor{Red}{\texttt{Late}}, \textcolor{Red}{\texttt{Late}} \texttt{-} \textcolor{Blue}{\texttt{NoFailure}}}                      \\ 

\multicolumn{1}{l|}{\texttt{ValueSubtle - ValueSubtle}} & \multicolumn{1}{l}{\textcolor{Blue}{\texttt{NoFailure}} \texttt{-} \textcolor{Red}{\texttt{Late}}, \textcolor{Blue}{\texttt{NoFailure}} \texttt{-} \textcolor{Blue}{\texttt{NoFailure}}, \textcolor{Red}{\texttt{Late}} \texttt{-} \textcolor{Red}{\texttt{Late}}, \textcolor{Red}{\texttt{Late}} \texttt{-} \textcolor{Blue}{\texttt{NoFailure}}}                      \\ 

\end{tabular}
\label{tab:test-results}
\vspace{-3mm}
\end{table*}

Notably, \texttt{Early} and \texttt{Late} failures are always propagated to output ports, whereas \texttt{ValueSubtle} and \texttt{ValueCoarse} produce different outcomes depending either on the position within the time series of the value affected by the mutation or the magnitude of the mutation. In fact, when a wrong voltage value, injected as either a \texttt{ValueSubtle} or a \texttt{ValueCoarse},  occurs at the beginning of the time series, changing the original value to a value near or below 0, the component responsible for turning on the water fans and the LEDs is markedly delayed, resulting in the detection of a \texttt{Late} failure on output ports. 
Instead, when the mutated voltage value is set close to 5 Volts, i.e., the maximum accepted value provided by the injector in input according to the use case, or even higher, no notable changes are detected, resulting in a \texttt{NoFailure}. This is because the Irrigation unit component in Proteus is configured to flatten voltage values up to 5 Volts. Interestingly, depending on the context of the failure and the specific value, it may either mask the failure or transform the failure into a failure of a different kind.

These results contributed to improving the knowledge of the engineers about the fault tolerance of the system. In fact, the engineers' supposed failures would only propagate unchanged through the component, while failure propagation rules show more complicated, sometimes context-dependent, patterns. Further, the component could sometime mask the effect of the same failures. 
%
%
For instance, by referring to Table \ref{tab:IUFLAtypes}, just one of the sample FLA rules (last row) was actually confirmed by our testing techniques. This is primarily because our testing approach has yet to cover the \texttt{Commission} and \texttt{Omission} failure types, which appear in the prior three rules in Table \ref{tab:IUFLAtypes}. Please note that not all the recommended tests listed in Table \ref{tab:test-results} will necessarily be used in the FLA analysis. However, having these results can provide the user with greater clarity regarding potential failure combinations, which can improve the accuracy and comprehensiveness of the FLA rules and generated fault trees.



\section{Related Work}
\label{sec:relatedWork}
In this section, we review  \textit{i)} the most relevant Model-Driven Engineering (MDE) approaches applied in the context of IoT \textit{ii) }applications of mutation testing in fault analysis, and, \textit{iii)} specification mining in software verification.

\subsection{MDE for \iot development}

 Ciccozzi \etal \cite{Ciccozzi2017} exploits the MDE paradigm to enable the abstraction of IoT systems. They propose exploiting the MDE paradigm to enable the abstraction of IoT systems, the easy handling of the various degrees of automation in software development, and the performance analysis of the system from different perspectives.
%

Thramboulidis \etal \cite{Thramboulidis2017} developed an MDE approach to face the complexity of IoT-based cyber-physical manufacturing systems. The conceived language allows domain experts to integrate IoT protocols during the system specification. 


 ThingML \cite{ThingMLcore} is an IoT engineering platform that combines well-proven textual software-modeling constructs aligned with UML, such as statecharts and components, with an imperative platform-independent action language for developing IoT applications. 
 

Fortas \etal \cite{Fortas2022} exploit MDE techniques to build an approach supporting the development and testing of IoT applications. In particular, they use ThingML \cite{ThingMLcore} in the modeling process and Proteus for simulation.



Monitor-IoT \etal \cite{MonitorIoT} is a graphical designer based on the Obeo Designer Community and Eclipse Sirius tools. The framework allows developers to model IoT multi-layer monitoring architectures. The tool enables the definition of computing nodes and their resources that support the monitoring processes, \ie data collection, transport, processing, and storage. Monitor-IoT is flexible enough to support the modeling at the edge, fog, and cloud layers. 

\subsection{Mutation testing in fault analysis}
Praphamontripong \etal \cite{Praphamontripong2010} present an approach to testing Web applications by applying mutation analysis to the connections among Web application software components. The authors showed the effectiveness of mutation analyses in creating tests supporting fault detection. The proposed type of analysis is able to discover also new mutation operators. 

Similarly, Moran \etal \cite{Moran2018} applied mutation testing in the context of mobile Android applications. Besides the application of empirically derived operations, the proposed tool supports the automation of the process of detecting potential mutant locations, generating mutants, and discovering new operations. 



Humbatova \etal \cite{Humbatova2021} present an approach for testing Deep Learning (DL) solutions. The authors extracted mutation operators from existing fault taxonomies. Then, they assessed the mutation operators to understand whether they produce killable, but not trivial, mutations. Eventually, they evaluated the approach by comparing it with the existing DeepMutation++ DL mutation tool. The results showed that their operators can discriminate more effectively between a weaker from a more robust test set.

Belli \etal \cite{BELLI201625} propose a model-based mutation testing approach for industrial systems based on directed graphs. The approach generates mutants and injects faults at the model level. In such a way, the mutation testing strategy can be applied even when the source code is unavailable. They only use two mutation operators, omission and insertion, by means of directed graphs. Then, these graphs are semantically enriched and exemplified using a collection of graph-based models to generate other operators.



\subsection{Specification Mining}

Concerning the understanding of system behaviour, \emph{specification mining}, intended as the extraction of high-level specifications from existing code, may play a key role. Approaches exploiting mined specifications can be used for program understanding but also for formal verification.

%

Dallmeier \etal \cite{Dallmeier2010} propose \emph{TAUTOKO}, a typestate 
 miner that combines systematic test case generation and typestate mining. Using those strategies. the approach systematically extends the execution space and enriches the final specification by increasing true positives. 


The need for good specifications for effective system verification is also highlighted by Cao \etal \cite{Cao2018RulebasedSM}. They adopted a rule-based specification mining approach that explores the
search space of all possible rules and uses interestingness measures to differentiate specifications from false positives. Then, the authors propose a learning-to-rank-based approach to consider 38 available interestingness measures together and investigate their combinations.
Their experiment results show that the learning-to-rank-based approach can improve the best ranking performance using a single measure by up to 66\%.

 \emph{ARTINALI++} \cite{ALIABADI2021111016} tool dynamically mines specifications of Complex Cyber-Physical Systems (CPS) to manage security issues. The approach generates a multi-dimensional model that is capable of embodying time, data, and events into the specifications. \emph{ARTINALI++} has been validated using three CPS platforms for intrusion detection. The results showed an average of 97.7\% detection accuracy across platforms while incurring reasonable performance and memory overheads.


\section{Conclusion and Future Work}
\label{sec:conclusion}
This paper discussed the challenges and importance of supporting early safety analysis of IoT systems, which are susceptible to various failures that can impact their functionality and the environment they operate in. Failure propagation within these systems is complex due to their diverse components and distributed nature. To address this, the paper discusses using Failure-Logic Analysis (FLA) to understand how component failures may propagate and affect the system's behavior. However, FLA relies on manually specified rules, which can be error-prone and incomplete. The paper proposes adopting testing methodologies to mitigate the issues with manually specified FLA rules. Potential faults can be observed and identified by subjecting the IoT system to various test cases. By means of the proposed testing techniques, it is possible to support the validation of the correctness of the system's behavior and the effectiveness of the specified rules in capturing fault scenarios. Future plans include the support of all the fault types that can be specified at the level of IoT system modeling. Moreover, we intend to investigate the generalizability of the proposed technique by considering different execution environments than Proteus.

\section*{Acknowledgment}
\addcontentsline{toc}{section}{Acknowledgment}
This work has been partially supported by the EMELIOT national research project, which has been funded by the MUR under the PRIN 2020 program (Contract 2020W3A5FY)

\bibliographystyle{IEEEtran}  
\bibliography{main}

\end{document}